\documentstyle[epsf,epsfig,referee]{mn}

\newcommand{\beq}{ \begin{equation} }
\newcommand{\eeq}{ \end{equation} }

\title{The ionization fraction in $\alpha$--models of protoplanetary
disks}
 
\author[Fromang, Terquem \& Balbus] {S\'ebastien Fromang$^1$, Caroline
Terquem$^{1,2}$ and Steven A. Balbus$^3$ \\ 
$^1$ Institut d'Astrophysique de Paris, 98 bis Boulevard Arago, 75014
Paris, France \\
$^2$ Universit\'e Denis Diderot--Paris VII, 2 Place Jussieu, 75251
Paris Cedex 5, France \\ 
$^3$ Virginia Institute of Theoretical Astronomy, Dept.~of Astronomy,
University of Virginia, Charlottesville, VA 22903-0818, USA}

\date{Accepted.
      Received;
      in original form }

\pubyear{}

\begin{document} 

\maketitle

\begin{abstract} 

We calculate the ionization fraction of protostellar $\alpha$ disks,
taking into account vertical temperature structure, and the possible
presence of trace metal atoms.  Both thermal and X--ray ionization are
considered.  Previous investigations of layered disks used radial
power--law models with isothermal vertical structure.  But $\alpha$
models are used to model accretion, and the present work is a step
towards a self-consistent treatment.  The extent of the magnetically
uncoupled (``dead'') zone depends sensitively on $\alpha$, on the
assumed accretion rate, and on the critical magnetic Reynolds number,
below which MHD turbulence cannot be self--sustained.  Its extent is
extremely model-dependent.  It is also shown that a tiny fraction of
the cosmic abundance of metal atoms can dramatically affect the
ionization balance.  Gravitational instabilities are an
unpromising source of transport, except in the early stages
of disk formation.

\end{abstract}  


\section{Introduction} 

The only process known that is able to initiate and sustain turbulent
transport in accretion disks is the magnetorotational instability
(MRI; Balbus \& Hawley 1991, 1998 and references therein).  Because
of the presence of finite resistivity, however, protostellar disk
applications of the MRI are not straightforward.  Numerical
magnetohydrodynamic (MHD) disk simulations with Ohmic dissipation
(Fleming et al. 2000) show that magnetic turbulence cannot be
sustained if the magnetic Reynolds number, $Re_M$, is lower than some
critical value, $Re_{M, {\rm crit}}$.  If there is a net magnetic flux
through the disk, the simulations indicate that $Re_{M, {\rm crit}}$
is about 100, and roughly corresponds to the Reynolds number below
which the modes are linearly stable.  If there is no net magnetic flux
through the disk, $Re_{M, {\rm crit}}$ is found to be much larger, on
the order of $10^4$.  In this case, MHD turbulence can be suppressed
even if the linear modes are only slightly affected.  It is important
to bear in mind that the value of $Re_{M, {\rm crit}}$ is uncertain.
Recent studies of the linear stability of protostellar disks indicate
that the effects of Hall electromotive forces are important, and that
the actual critical Reynolds number may accordingly be smaller (Wardle
1999, Balbus \& Terquem 2001).

On scales of 1~AU in a protostellar disk, $Re_M=100$ corresponds to
ionization fraction of about $10^{-12}$ (e.g., Balbus \& Hawley 2000
and \S~\ref{sec:results} below).  Although very small, this fraction
may not be attained in the intermediate regions of protostellar disks.
The disk zones in which $Re_M$ is lower or larger than the critical
Reynolds number are usually referred to as {\em dead} and {\em
active}, respectively (Gammie 1996).  The active layer extends
vertically from the disk surface down to some altitude, which depends
upon the radial location and the disk model.

The extent of the dead zone has been modeled by several authors using
different ionization agents: Gammie (1996, cosmic rays), Igea \&
Glassgold (1999, X-rays) and Sano et al. (2000, cosmic rays and
radioactivity).  All these studies were based on the 
minimum mass disk model of
Hayashi, Nakazawa \& Nakagawa (1985), in which temperature and surface
mass density vary as a simple power law of the radius, and the
vertical structure is isothermal.  This is a rather arbitrary choice,
and the large scale structure of a turbulent $\alpha$~disk is in fact
very different (Papaloizou \& Terquem 1999).  There is some
theoretical evidence that MHD turbulence leads to a large scale
$\alpha$~type structure in thin Keplerian disks (Balbus \& Papaloizou
1999).  It is the goal of this paper to investigate the ionization
fraction of an $\alpha$--type disk, taking into account the
vertical structure of such models.  We shall consider thermal ionization
and X--ray ionization, as these mechanisms are likely to be more
important than cosmic rays in X-ray active young stellar objects
(Glassgold, Feigelson \& Montmerle 2000).

The disk ionization depends on the recombination rate of the
electrons, which are removed through dissociative recombination with
molecular ions and, at a much slower rate, through radiative
recombination with heavy metal ions.  In the previous studies of disk
ionization by X--rays, it was assumed that all the metal atoms were
locked up in dust grains most of which had themselves sedimented
toward the disk midplane.  However, the ionization fraction is
extremely sensitive to even a very small number of metal atoms.  This
is because these rapidly pick up the charges of molecular ions and
recombine only slowly with the electrons.  In this paper, we therefore
include the effect of a non zero density of metal atoms on the
disk ionization.

The plan of the paper is as follows: In \S~\ref{sec:models}, we
describe the disk models used, and compare them with Hayashi et
al. (1985).  In \S~\ref{sec:ionization}, we discuss the different
ionization mechanisms.  Thermal ionization is important in the disk
inner parts, whereas X--ray ionization dominates everywhere else.  We
also discuss the effect of the presence of heavy metal atoms on the
ionization fraction.  In \S~\ref{sec:results} we present results for
$\alpha$--type disks for a range of gas accretion rate $\dot{M}$ and
values of the viscosity parameter $\alpha$.  We find that the extent
of the dead zone depends very sensitively on the critical Reynolds
number, the parameters of the disk model ($\dot{M}$ and $\alpha$) and
the density of heavy metal atoms.  With no metal atoms and $Re_{M,
{\rm crit}}=100$, we find in most cases that the dead zone generally
extends from a fraction of an AU out to $10$ to $10^{2}$~AU.  With an
accretion rate of $\dot{M}=10^{-8}$~M$_\odot$~yr$^{-1}$ and
$\alpha=10^{-2}$ for instance, the dead zone extends from 0.2 to
100~AU.  This is much larger than what was found by previous authors,
who used a smaller value of $Re_{M, {\rm crit}}$.  However, we also
find that the dead zone disappears completely for $\alpha \ge 10^{-2}$
when there is even a tiny density of heavy metal atoms.  For instance,
a density as small as $10^{-7}$ or $10^{-6}$ times the cosmic
abundance is enough to make a disk with
$\dot{M}=10^{-8}$~M$_\odot$~yr$^{-1}$ and $\alpha=10^{-2}$ completely
turbulent.  The dead zone is dramatically reduced or even disappears
when the critical Reynolds number is taken to be 1, even when there
are no heavy metal atoms.  In \S~\ref{sec:evolution}, we study the
evolution of an $\alpha$ disk with a dead zone, with the aim of
investigating local gravitational instability.  Gammie (1996, 1999)
and Armitage, Livio \& Pringle (2001) have noted that a layered disk
cannot accrete mass steadily, and that accumulation of mass in the
dead zone may lead to gravitational instabilities.  We find that when
there is no mass falling onto the disk, the accumulation rate is too
slow for gravitational instabilities to develop within the disk
lifetime.  Finally, in \S~\ref{sec:discussion} we summarize and
discuss our results.

\section{Disk models}
\label{sec:models}

\begin{figure}
\centerline{
\epsfig{file=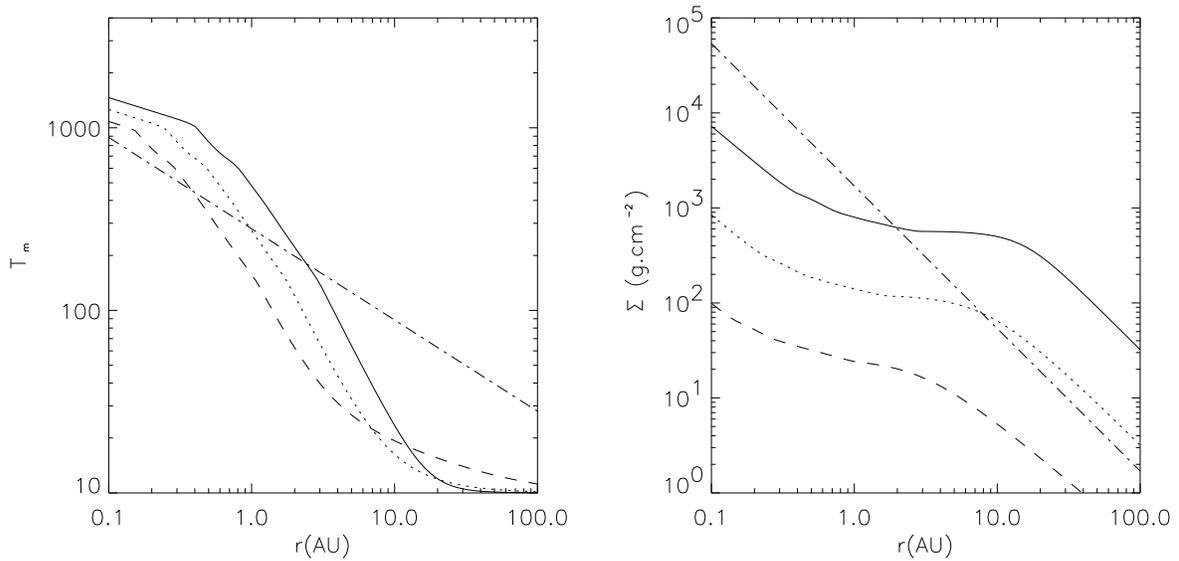,width=16.cm} }
\caption[] {Shown is the midplane temperature $T_m$ in K ({\em left
panel}) and the surface mass density $\Sigma$ in g~cm$^{-2}$ ({\em
right panel}) vs. $r$ in AU for $\dot{M}=10^{-8}$~M$_\odot$~yr$^{-1}$
and $\alpha=10^{-3}$ ({\em solid lines}), $10^{-2}$ ({\em dotted
lines}) and 0.1 ({\em dashed lines}).  For comparison, the model used
by Igea \& Glassgold (1997) is also displayed ({\em dotted--dashed
lines}).}
\label{fig0}
\end{figure}

We use conventional $\alpha$--disk models, as calculated by
Papaloizou \& Terquem (1999), to which the reader is referred for
details.  The disk is assumed to be in Keplerian rotation around a
central mass $M_{\ast}=1$~M$_{\odot}$.  The opacity, taken from Bell
\& Lin (1994), has contributions from dust grains, molecules, atoms
and ions.   The values of $\alpha$ and mass flow rate $\dot{M}$
are taken to be free parameters, and determine the model uniquely.
In the steady state limit, $\dot{M}$ is constant through the disk.
At a given radius $r$, the vertical structure is obtained by solving
the equations of hydrostatic equilibrium, energy conservation and
radiative transport with appropriate boundary conditions.  (At the
temperatures of interest here, convective transport is not
significant).  The detailed results of these calculations for
$\alpha=10^{-2}$ or 0.1 and $\dot{M}$ in the range
$10^{-9}$--$10^{-6}$~M$_{\odot}$~yr$^{-1}$ may be found in Papaloizou
\& Terquem (1999).  In Figure~\ref{fig0}, we plot steady state values
of the midplane temperature $T_m$ and the surface mass density $\Sigma$
versus $r$ for $\alpha$ between $10^{-3}$ and $10^{-1}$; for
illustrative purposes we have adopted
$\dot{M}=10^{-8}$~M$_\odot$~yr$^{-1}$.

All solar nebula ionization studies previous to this have used the
Hayashi et al.~(1985) model of the solar nebula:
\begin{displaymath} 
\Sigma = 1700 \left( \frac{1 \; {\rm AU}}{r} \right)^{1.5} \; {\rm
g~cm}^{-2} \; \; {\rm and} \; \; 
T = 280 \left( \frac{1 \; {\rm AU}}{r} \right)^{0.5} \; {\rm K} ,
\end{displaymath} 

\noindent as the equilibrium profile.  (Here, $r$ is the cylindrical
radius.) This model is based upon chain of arguments: the current
orbits and composition of the planets reflect the distribution and
composition of dust in the preplanetary nebula; the planets have not
moved radially in the course of their history; the formation of
planets was extremely efficient.  While this has been an useful
organization of a complex problem, it certainly leaves room for other
approaches to modeling the nebula.  A question of some interest is how
important the nebular model is to the ionization structure, which has
motivated the approach presented here.  For example, the Hayashi et
al. (1985) model leads to a more centrally condensed disk mass than
that obtained in a steady $\alpha$--disk.  With $\alpha=10^{-2}$, a
very large accretion rate of $\dot{M} \sim
10^{-6}$~M$_{\odot}$~yr$^{-1}$ is necessary to obtain the above
Hayashi value of $\Sigma$ at 1~AU.  For comparison, we have plotted
temperature and mass density curves of the two models in
Figure~\ref{fig0}.

Ionization by X--rays is thus much more efficient around 1~AU in an
$\alpha$--disk model than in the Hayashi et al.  (1985) model.  The
theoretical justification for a minimum mass model is not strongly
compelling, and it is quite incompatible with standard accretion disk
theory.  In addition, as mentioned above, there is some theoretical
evidence that MHD turbulence leads to a large scale $\alpha$~type
structure in thin Keplerian disks (Balbus \& Papaloizou 1999).  This
is the primary motivation for the work presented here.

\section{Ionization and Recombination}
\label{sec:ionization}

Protostellar disks are ionized mainly by thermal processes and by
nonthermal X--rays.  Cosmic rays, a classical ionization source in
cool gas, have also been considered as a source of ionization (Gammie
1996, Sano et al. 2000), but the low energy particles (important for
ionization) were almost certainly excluded by winds from the early
solar nebula, as they are today in a far less active environment.
Radioactive decay of ${}^{40}$K and ${}^{26}$Al has also been
investigated by some authors (e.g., Consolmagno \& Jokipii 1978), but
their ionization effects are quite small compared to the levels of
interest here (Stepinski 1992, Gammie 1996).

\subsection{Thermal ionization}
If not condensed onto grains, the alkali ions Na$^+$ and K$^+$
will be the dominant thermal ionization source
in protostellar disks (Umebayashi \& Nakano 1983). 
At the onset of dynamically interesting ionization levels ($\sim
10^{-13}$), the K$^+$ ion is, with its smaller ionization potential,
more important.  In this regime, the Saha equation may be approximated
as (Balbus \& Hawley 2000):

\begin{equation}
x_{e} \equiv \frac{n_e}{n_n} =
6.47 \times 10^{-13} 
\left( \frac{a}{10^{-7}} \right)^{1/2}
\left( \frac{T}{10^3} \right)^{3/4}
\left( \frac{2.4 \times 10^{15}}{n_n} \right)^{1/2}
\frac{{\rm exp} \left( -25188/T\right)}{1.15 \times 10^{-11}}
\label{saha}
\end{equation}

\noindent where $n_e$ and $n_n$ are respectively the electron and
neutral number densities in cm$^{-3}$, and $a$ is the K abundance
relative to hydrogen.  Due to the Boltzmann cut--off factor, thermal
ionization is important only in the disk inner regions, on scales less
than an AU, where the midplane temperature is likely to exceed
$10^3$~K.  (Above this temperature the alkalis will tend to be in the
gas phase, making the approximation self-consistent.)  If there is
magnetic coupling on scales larger than this, nonthermal ionization
sources are required.

\subsection{X--ray ionization}

Young stellar objects appear to be very active X-ray sources, with
X-ray luminosities in the range of $10^{29}$--$10^{32}$~erg~s$^{-1}$
and photon energies from about 1 to 5~keV (Koyama et al.~1994,
Casanova et al.~1995, Carkner et al.~1996).  Glassgold, Najita \&
Igea (1997, hereafter GNI97, see also Igea \& Glassgold 1999) pointed
out that these X-rays are likely to be the dominant nonthermal
ionization source in protostellar disks.  These authors modeled the
X-ray source as an isothermal ($T=T_X$) bremsstrahlung coronal ring,
of radius of about $10~R_{\odot}$, located at a similar distance above
(and below) the disk midplane.  The total X-ray luminosity is $L_X$,
with each hemisphere contributing $L_X/2$.
The associated ionization rate is given by (Krolik \& Kallman
1983, GNI97):
\begin{equation}
\zeta = \frac{(L_X/2)}{4\pi r^2 kT_X} \; \sigma \left( kT_X
\right) \; \frac{kT_X}{\Delta \epsilon} \; J(\tau).
\label{xray}
\end{equation}
\noindent Here, $\sigma$ is the photoionization cross section,
which is fit to a power law (Igea \& Glassgold 1999):
\begin{equation}
\sigma(E) = 8.5\times 10^{-23} (E/{\rm keV})^{-n},
\end{equation}
with $n=2.81$ (these values apply to the case where heavy elements are
depleted onto grains and get segregated from the gas).  $\Delta
\epsilon=37$~eV is the average energy required by a primary
photoelectron to make a secondary ionization.  The dimensionless
integral $J$ is \beq J(\tau) = \int^\infty_{x_0} x^{-n} \exp(-x -\tau
x^{-n} )\, dx , \eeq an energy integral over the X-ray spectrum
involving the product of the cross section (whence the factor
$x^{-n}$) and the attenuated X-ray flux.  The factor $\tau$ is the
optical depth at an energy of $kT_X$; it depends upon one's location
within the disk.  Generally the integral is insensitive to the lower
limit threshold energy represented by $x_0$, and for the large $\tau$
case of interest here, it may be asymptotically expanded.  The leading
order result is \beq J(\tau) \simeq A \tau^{-a} \exp(-B\tau^{-b}) \eeq
where $A=0.686$, $B= 1.778$, $a= 0.606$, $b = 0.262$.
In computing the optical depth,
we make the approximation that the photons travel
along straight lines; that is, we will neglect their diffusion both by
the ambient medium and by the disk interior.  Note however that
scattering by the disk atmosphere would increase the ionization, both
because some of the photons directed away from the disk would be
scattered back, and because scattering provides
pathways to the disk interior with smaller optical depths than a simple
linear traversal.  Note that here we do not make the approximation of
GNI97 that the path of the photons inside the disk is vertical.

\subsection{Recombination processes}

In contrast to ionization, which is reasonably straightforward,
electron recombination is greatly complicated by the presence of dust
grains in the nebula.  Not only are results dependent upon the size
spectrum of the grains, imperfectly understood surface physicochemical
processes will strongly influence charge capture and emission.
Uncertainties attending the role of dust grains represent the greatest
obstacle in estimating the solar nebula's ionization structure.

Gammie (1996) and GNI97 finessed this issue by arguing that the dust
will settle rapidly toward the midplane, and that the dominant
recombination process will therefore be molecular dissociative
recombination.  While this greatly simplifies matters, it is prudent to
regard the vertical distribution of the grains in magnetically coupled
disk regions as an outstanding problem.  Turbulence tends to mix,
but the MRI is not particularly efficient at mixing {\em vertically.}
Convective turbulence, should it be present, is a fine vertical mixer,
and dust emission in the upper layers enhances the cooling, aiding the
convection process itself.  But sustaining convection without the MRI
is known to be problematic.

Given the uncertainties, we take the view that it makes little sense
to strive for high accuracy in a complex model.  We have therefore
opted for the simplicity of the previous studies, and will restrict
our modeling to gas phase recombination.  This may well overestimate
the electron density in the regime in which adsorption by grains
exceeds X-ray photo-emission, but the model is readily understood and
serves as a useful benchmark.  It is technically valid when the grains
lie within a dead (magnetically decoupled) layer near the
midplane, or if vertical mixing is inefficient.  In the latter case,
note that the disk is likely to develop a dead sheet at its
midplane where the dust has concentrated, as in either the Gammie
(1996) or GNI97 models.

One important addition we shall emphasize is the presence of charge
exchange between molecular ions and metal atoms (Sano et al. 2000).
What is of interest here is the great sensitivity of $x_e$ to the
presence of even trace amounts of metal atoms, and the extreme {\em
insensitivity} of the metal atoms to the spatial density of the
grains.  The latter is a consequence of the dual role played by the
dust: they are adsorbers as well as (X-ray induced) desorbers
of metal atoms.  

If thermal processes prevailed in the disk, the metal atom population
would be negligibly small even by these sensitive standards: the
temperatures of interest are well below the condensation temperatures
of most refractory elements.  But just as nonthermal processes may regulate the
ionization fraction of the disk, they may also regulate metal atom
population.  Consider a dust grain in the presence of an X-ray
radiation field.\footnote{Cosmic rays, whose ionization
effects are secondary to X-rays, may in fact be more effective
at releasing metal atoms from grains.  Their presence, while increasing
the rate of sputtering, would not effect the qualitative point being
made.}  The rate at which atoms are liberated from the grain
is ${\cal F}\sigma_{rad} y$, where ${\cal F}$ is the photon number
flux, $\sigma_{rad}$ is the radiation cross section, and $y$ is the
quantum yield.  This is balanced by the metal adsorption rate onto the
grain, $n_M v_{th} \sigma_{cap} s$, where $n_M$ is the metal atom
density, $v_{th}$ is the thermal velocity, $\sigma_{cap}$ is the
capture cross section, and $s$ is the sticking probability.  Thus,

\beq\label{nM} x_M\equiv\frac{n_M}{n_n} = \frac{{\cal F}} {n_n v_{th}}
\frac {\sigma_{rad}}{\sigma_{cap}} \frac{y}{s}, \eeq 

\noindent independent of the dust density.  To understand the
consequences of this more clearly, we next examine the dependence of
$x_e$ upon $x_M$.

The ionization fraction $x_e$ is obtained by balancing the net rates of
ionization and recombination.  Electrons are captured via dissociative
recombination with molecular ions (e.g., HCO$^+$) and radiative
recombination with heavy metal ions (e.g., Mg$^+$).  As noted, charges
are also transferred from molecular ions to metal atoms, hence the
level of ionization depends on the abundance of metal atoms in the
gas.

Let $n_m$, $n_{m^+}$, and $n_{M^+}$, respectively denote the density
of molecules, molecular ions, and metal ions.  The rate equations for
$n_e$ and $n_{m^+}$ are:

\begin{equation}
\frac{dn_e}{dt} = \zeta n_n - \beta n_{m^+} n_e - \beta_r n_{M^+} n_e ,
\label{dne}
\end{equation}

\begin{equation}
\frac{dn_{m^+}}{dt} = \zeta n_n - \beta n_e n_{m^+} - \beta_t n_M n_{m^+} ,
\label{dnm}
\end{equation}

\noindent where $\beta$ is the dissociative recombination rate
coefficient for molecular ions, $\beta_r$ the radiative recombination
rate coefficient for metal atoms, and $\beta_t$ the rate coefficient of
charge transfer from molecular ions to metal atoms.  We have made the
standard simplifying assumption that the rate coefficients are the same
for all the species.  For the numerical work, we take (Oppenheimer \&
Dalgarno 1974; Spitzer 1978; Millar, Farquahr, \& Willacy 1997 for
$\beta_t$):
\beq\label{betas}
\beta_r  =  3 \times 10^{-11} \; T^{-1/2} \; \; {\rm cm}^3,
\quad
\beta  = 3 \times 10^{-6} \; T^{-1/2} \; \; {\rm cm}^3,
\quad
\beta_t  =  3 \times 10^{-9} \; \; {\rm cm}^3 \; {\rm s}^{-1}.
\eeq

\noindent Finally, charge neutrality implies

\begin{equation}
n_e = n_{M^+} + n_{m^+} .
\label{ne}
\end{equation}

In steady-state equilibrium, equations~(\ref{dne}), (\ref{dnm}) and
(\ref{ne}) lead to (Oppenheimer \& Dalgarno 1974):

\begin{equation}\label{cubic}
x_e^3 + \frac{\beta_t}{\beta} x_M x_e^2 - \frac{\zeta}{\beta n_n} x_e
- \frac{\zeta \beta_t}{\beta \beta_r n_n} x_M = 0,
\end{equation}

\noindent where $x_M \equiv n_M / n_n$.  The extreme sensitivity
of this equation to $x_M$ is apparent upon substituting from
equation (\ref {betas}) for $\beta_t/\beta$ and $\beta_t/\beta_r$:

\beq\label{cubbis}
x_e^3 + 10^{-3} T^{1/2}  x_M x_e^2 - \frac{\zeta}{\beta n_n} x_e
- 10^2 T^{1/2} \frac{\zeta}{ \beta n_n} x_M = 0.
\end{equation}

In the absence of metals ($x_M = 0$), equation (\ref{cubic}) has the
simple solution used by Gammie (1996) and GNI97:

\begin{equation}
x_e = \sqrt{\frac{\zeta}{\beta n_n}},
\label{xM0}
\end{equation}

\noindent
a balance between the first and third terms on the left side of the
equation.  This case would correspond, for example, to all metals
locked in sedimented grains.  The opposite limit, metal domination, is
equally simple:

\begin{equation}
x_e = \sqrt{\frac{\zeta}{\beta_r n_n}},
\end{equation}

\noindent
and corresponds to a balance between the second and fourth terms of
the cubic.  The transition from one limit to the other begins when the
last two terms of the left--hand--side of the equation are comparable,
that is when \beq x_M \sim 10^{-2} T^{-1/2} x_e, \eeq from which one
immediately sees the extreme sensitivity to the metal atom abundance.
A value of $x_M \sim 10^{-14}$, or about $10^{-7}$ of its cosmic
abundance, might well affect the ionization balance.  This, combined
with insensitivity of low values of $x_M$ to the dust abundance, leads
us to consider the effect of finite $x_M$ on the ionization balance in
the disk.

\section{Ionization fraction in $\alpha$--disk models} 
\label{sec:results}

For a given disk model with fixed $\dot{M}$ and $\alpha$, we calculate
the ionization fraction $x_e$ as a function of $r$ and $z$ using
equations~(\ref{saha}) and~(\ref{cubic}), i.e. taking into account
both thermal and X--ray ionization.  In equation~(\ref{cubic}), the
fraction of heavy metal atoms, $x_M$, is varied between zero and some
finite value, in order to determine the minimum value of $x_M$ for
which a particular disk model is sufficiently ionized to be
magnetically coupled.  In \S~\ref{sec:discussion}, we show that the
values of $x_M$ we have used are at least roughly consistent with
equation~(\ref{nM}).

The Ohmic resistivity $\eta$ is (e.g., Blaes \& Balbus 1994):

\begin{equation}
\eta = \frac{234}{x_e} \; T^{1/2} \; {\rm{cm \; s^{-2}}},
\end{equation}

\noindent and we define the magnetic Reynolds number as:

\begin{equation}
Re_M = \frac{c_s H}{\eta},
\end{equation}

\noindent where $H$ is the disk semithickness and $c_s$ is the sound
speed.  Numerical simulations including Ohmic dissipation (Fleming et
al. 2000) indicate that MHD turbulence cannot be sustained if $Re_M$
falls below a critical value, $Re_{M, {\rm crit}}$, which depends upon
the field geometry.  If there is a mean vertical field present, $Re_{M,
{\rm crit}}$ is about 100.  However, recent studies of the linear
stability of protostellar disks including the effects of Hall
electromotive forces suggest that $Re_{M, {\rm crit}}$ may be smaller
(Wardle 1999, Balbus \& Terquem 2001), and preliminary results from
nonlinear Hall simulations appear to back this finding (Sano \& Stone, private
communication).  In anticipation of this, it is prudent to consider both $Re_{M,
{\rm crit}}=100$ and $Re_{M, {\rm crit}}=1$.

\begin{figure}
\centerline{
\epsfig{file=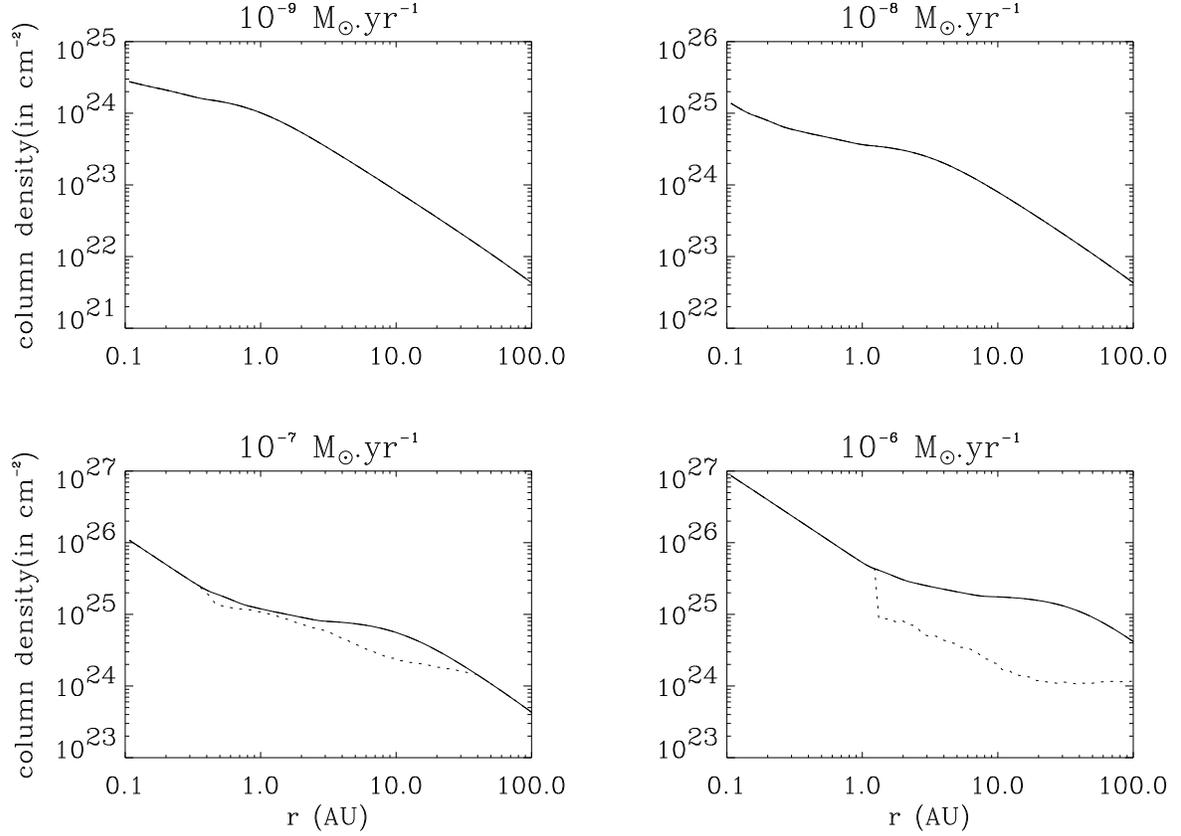,width=16.cm, angle=0} }
\caption[] {Column density, in cm$^{-2}$, of the whole disk ({\em
solid lines}) and of the active zone for $Re_{M, {\rm crit}}=100$ ({\em dotted
lines}). 
Here $x_M=0$, $\alpha=0.1$ and $\dot{M} = 10^{-9}$ ({\em upper left
panel}), $10^{-8}$ ({\em upper right panel}), $10^{-7}$ ({\em lower
left panel}) and $10^{-6}$~M$_\odot$~yr$^{-1}$ ({\em lower right
panel}).  When the curve representing the active zone coincides with
that representing the whole disk, then the whole disk is active. This
is the case for $\dot{M} = 10^{-9}$ and $10^{-8}$~M$_\odot$~yr$^{-1}$
when $Re_{M, {\rm crit}}=100$, and for all the values of $\dot{M}$
when $Re_{M, {\rm crit}}=1$. }
\label{fig1}
\end{figure}

\begin{figure}
\centerline{
\epsfig{file=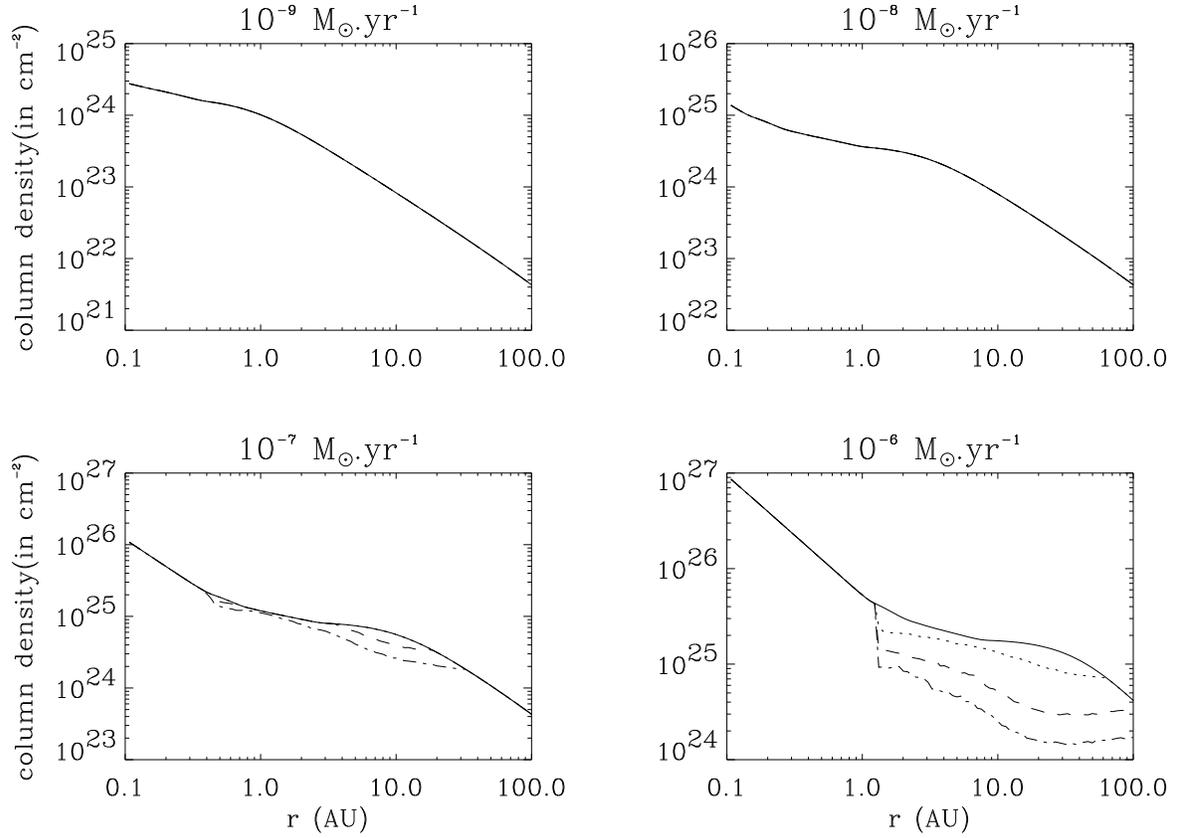,width=16.cm, angle=0}
}
\caption[] {Same as Fig.~\ref{fig1} but for $x_M \ne 0$.  On each
panel, the column density of the active zone is represented for
$x_M=10^{-16}$ ({\em dotted-dashed lines}), $10^{-15}$ ({\em dashed
lines}) and $10^{-14}$ ({\em dotted lines}).  For $\dot{M} = 10^{-9}$
and $10^{-8}$~M$_\odot$~yr$^{-1}$, the whole disk is active for all
these values of $x_M$.  For $\dot{M} = 10^{-7}$~M$_\odot$~yr$^{-1}$,
the whole disk is active for $x_M > 10^{-15}$.}
\label{fig2}
\end{figure}

\begin{figure}
\centerline{
\epsfig{file=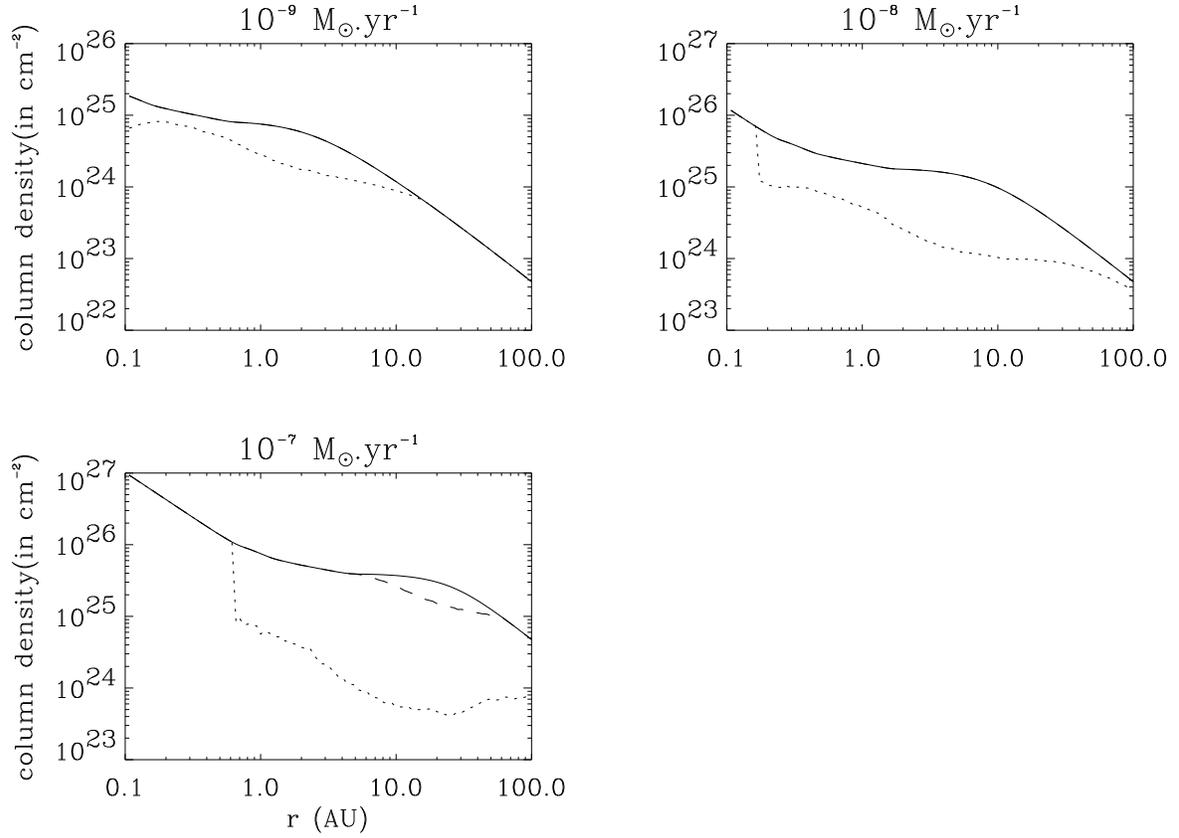,width=16.cm, angle=0} }
\caption[] {Same as Fig.~\ref{fig1} but for $\alpha=10^{-2}$ and for
both $Re_{M, {\rm crit}}=100$ ({\em dotted lines}) and $Re_{M, {\rm
crit}}=1$ ({\em dashed lines}).  The case corresponding to $\dot{M} =
10^{-6}$~M$_\odot$~yr$^{-1}$ has not been included as it gives a disk
mass unrealistically large.  Here, the whole disk is active when
$Re_{M, {\rm crit}}=1$ only for $\dot{M} = 10^{-9}$ and
$10^{-8}$~M$_\odot$~yr$^{-1}$.  }
\label{fig3}
\end{figure}

\begin{figure}
\centerline{ \epsfig{file=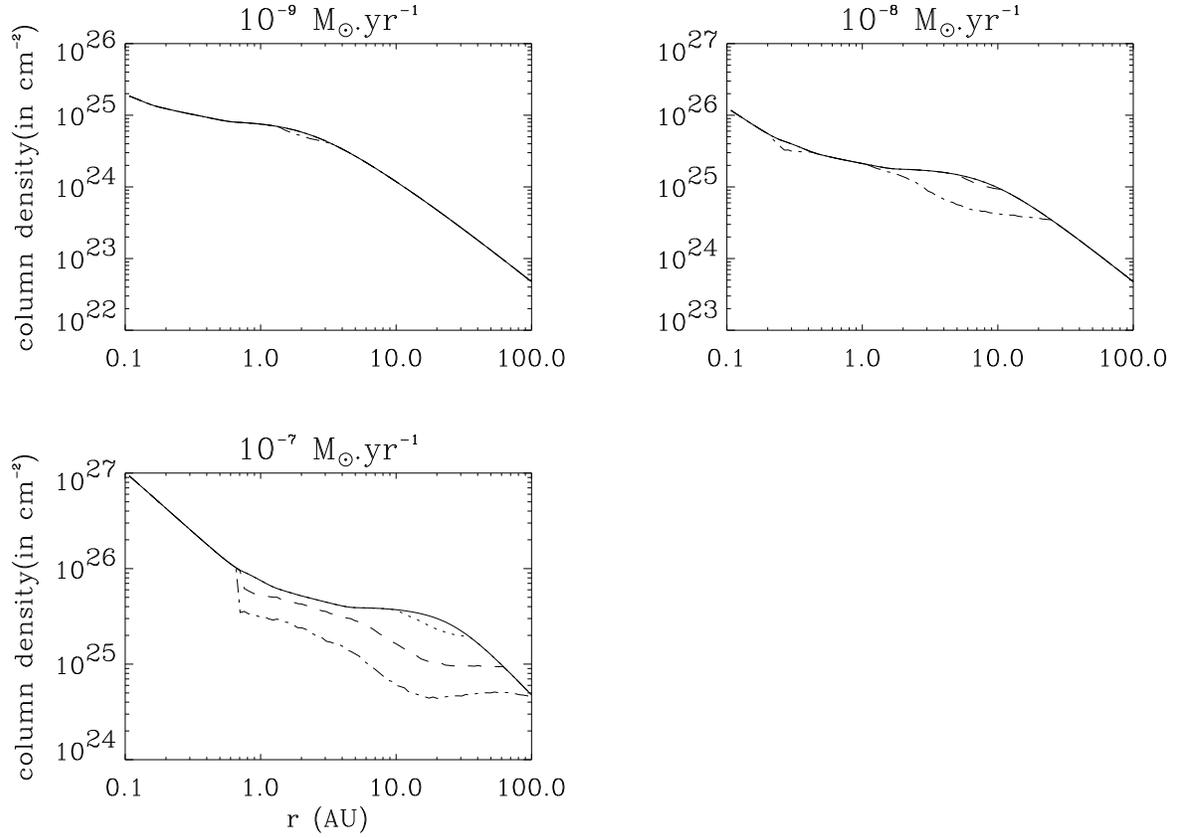,width=16.cm, angle=0}
}
\caption[] {Same as Fig.~\ref{fig2} ($Re_{M, {\rm crit}}=100$) but for
$\alpha=10^{-2}$ and for different values of $x_M$.  On each panel,
the column density of the active zone is represented for
$x_M=10^{-13}$ ({\em dotted-dashed lines}), $10^{-12}$ ({\em dashed
lines}) and $10^{-11}$ ({\em dotted lines}). The case corresponding to
$\dot{M} = 10^{-6}$~M$_\odot$~yr$^{-1}$ has not been included as it
gives a disk mass unrealistically large.  For $\dot{M} = 10^{-9}$,
$10^{-8}$ and $10^{-7}$~M$_\odot$~yr$^{-1}$, the whole disk is active
for $x_M > 10^{-13}$, $x_M > 10^{-12}$ and $x_M > 10^{-11}$,
respectively.}
\label{fig4}
\end{figure}

\begin{figure}
\centerline{ \epsfig{file=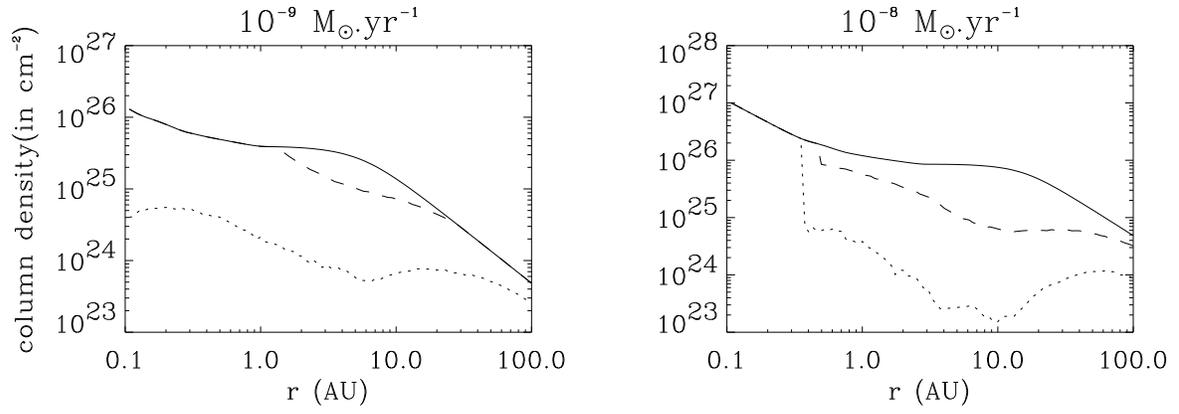,width=16.cm,angle=0} }
\caption[] {Same as Fig.~\ref{fig3} but for $\alpha=10^{-3}$.  Cases
with $\dot{M} = 10^{-7}$ and $10^{-6}$~M$_\odot$~yr$^{-1}$ have not
been included as they give a disk mass unrealistically large. }
\label{fig5} 
\end{figure}

\begin{figure}
\centerline{ \epsfig{file=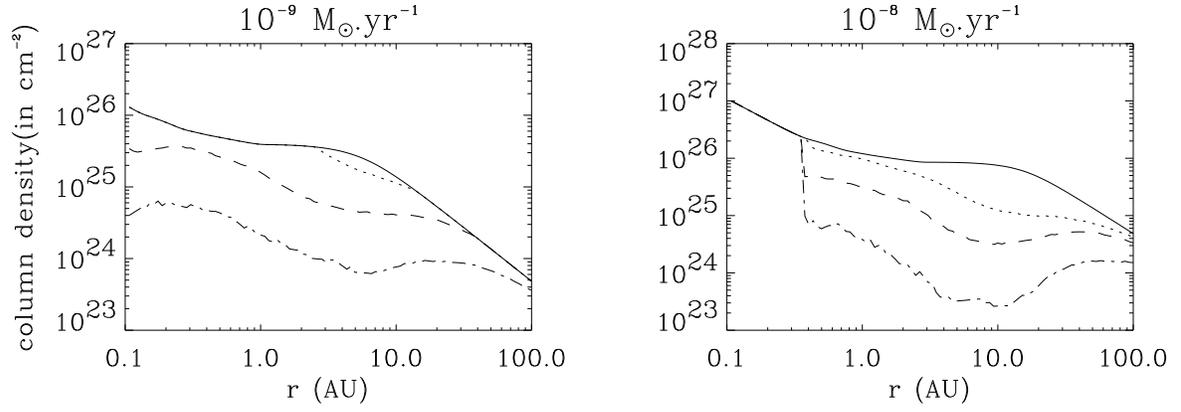,width=16.cm,angle=0}
}
\caption[] {Same as Fig.~\ref{fig2} ($Re_{M, {\rm crit}}=100$) but for
$\alpha=10^{-3}$ and for different values of $x_M$.  On each panel,
the column density of the active zone is represented for
$x_M=10^{-15}$ ({\em dotted-dashed lines}), $10^{-12}$ ({\em dashed
lines}) and $10^{-9}$ ({\em dotted lines}).  Cases with $\dot{M} =
10^{-7}$ and $10^{-6}$~M$_\odot$~yr$^{-1}$ have not been included as
they give a disk mass unrealistically large.  Here the dead zone never
disappears even for large values of $x_M$, and its extent does not
vary with $x_M$ once $x_M \ge 10^{-9}$.}
\label{fig6}
\end{figure}

In Figures~\ref{fig1}--\ref{fig5}, we plot the total column density,
and that of the active layer, for different disk models:
$\alpha=0.1$ (fig.~\ref{fig1} and~\ref{fig2}), $10^{-2}$
(fig.~\ref{fig3} and~\ref{fig4}) and $10^{-3}$ (fig.~\ref{fig5}
and~\ref{fig6}): $\dot{M}$ varies between
$10^{-9}$--$10^{-6}$~M$_\odot$~yr$^{-1}$.  Cases with $\alpha=10^{-2}$
and $\dot{M}=10^{-6}$~M$_\odot$~yr$^{-1}$ and $\alpha=10^{-3}$ and
$\dot{M} \ge 10^{-7}$~M$_\odot$~yr$^{-1}$ have not been considered, as
they give disk masses larger than that of the central star.  Note that
$x_M=0$ in figures~\ref{fig1}, \ref{fig3} and~\ref{fig5}, whereas $x_M$
is finite in figures~\ref{fig2} and~\ref{fig4}.  In all figures, the
X--ray luminosity and temperature are $L_X=10^{30}$~erg~s$^{-1}$ and
$kT_X=3$~keV respectively.  When the entire disk is found to be active,
a single curve is shown; otherwise, the difference between the two
curves indicates the column density of the dead layer.

In table~\ref{table1}, we summarize the dead zone properties for the
different models considered.  The quantity $x_{M,{\rm min}}$ is the
value of $x_M$ above which the dead zone disappears, and $r_{max}$ is
the radius at which the vertical column density of the dead zone is
largest.  The final column is the percentage of the vertical column
density occupied by the dead zone at $r=r_{max}$.  Note that if
$Re_{M, crit} = 1$, $\alpha = 0.1$ disks are all
active throughout their entire extent, as are $\alpha = 10^{-2}$
disks with $\dot{M} \le 10^{-8}$~M$_\odot$~yr$^{-1}$.  If one raises
$Re_{M, crit} $ to 100, $\alpha = 0.1$ disks are still magnetically active 
everywhere, when $\dot{M} \le 10^{-8}$~M$_\odot$~yr$^{-1}$.  As $\alpha$
increases, the disk surface density decreases at fixed $\dot{M}$, and the
active zone increases in vertical extent.  

Figures~\ref{fig2}, \ref{fig4}, and~\ref{fig6} show that as $x_M$
increases, the dead zone shrinks both radially (with its outer edge
moving inwards), and vertically.  Its inner edge does not move outwards,
because
it is located at the radius at which the temperature drops below the
level needed for thermal ionization to be effective, which is 
independent of $x_M$.  For a given $\alpha$, this radius increases
with $\dot{M}$, as the disk becomes hotter.  Similarly, this
radius increases when $\alpha$ decreases for $\dot{M}$ fixed.

The existence of a local maximum in the column density of the active
layer at $r=r_{max}$ is easily understood.  The $\alpha$--disk models
have an almost uniform surface mass density over a broad radial range
that depends on $\dot{M}$ and $\alpha$ (see fig.~\ref{fig0}).  As we
move away from the star remaining in this region, the X--ray flux
decreases because of simple geometrical dilution.  Since the column
density along the path of the photons stay about the same, the active
layer becomes thinner.  It starts to thicken only when the surface
mass density in the disk begins to drop.

Table~\ref{table1} gives the value of $x_M$ above which the disk is
completely active.  This should be compared with the cosmic abundance
of the metal atoms present in a disk, which is $2 \times 10^{-6}$ for
Na and $3 \times 10^{-5}$ for Mg (Anders \& Grevesse 1989, Boss 1996).
When $Re_{M, {\rm crit}}=100$ and for a disk model with
$\alpha=10^{-2}$ and $\dot{M}=10^{-8}$~M$_\odot$~yr$^{-1}$, we see
that an abundance of only $10^{-6}$ or even $10^{-7}$ of cosmic
(depending on the species) is needed for the disk to be active.

When $\alpha \le 10^{-3}$, the disk is never fully active, even for
large values of $x_M$.  This is because the disk is now very massive
(about 0.3 solar mass within 100~AU for
$\dot{M}=10^{-8}$~M$_\odot$~yr$^{-1}$), and there is a zone in the
disk where X--rays simply cannot penetrate.  Increasing $x_M$ clearly
cannot prevent this!

\section{Evolution of a disk with a dead zone}
\label{sec:evolution}

The fate of the magnetically coupled upper disk layers is
far from clear.  This relatively low density region may
emerge in the form of a disk wind; it is also possible that
the layers will accrete.  This is the assumption of Gammie (1996),
but it awaits verification by MHD simulations of stratified, partially
ionized disks.  

Should it occur, layered accretion in a disk will not be steady (Gammie
1996).  The regions of the disk where there is an inactive layer act
like a bottleneck through which the accretion is slowed and
where matter therefore accumulates.  Here we investigate whether enough
mass can accumulate to make the disk locally gravitationally unstable.
Of course, our calculation is in some sense not self-consistent, since
it is based upon the assumption that the disk structure is a constant
$\alpha$ model which we then show cannot be maintained! 
However, our
interest is limited to estimating the extent to which upper layer accretion can
build up the column density of the inner disk to the point of
gravitational collapse.  For this purpose,
the precise vertical structure is likely to
be a correction of detail, not a fundamental change.

Armitage et al. (2001) have considered a disk 
embedded in a collapsing envelope, so that mass is
continuously added to the disk at a rate
larger than the accretion through the disk itself.  This 
strongly favors the development of gravitational instabilities.  Here,
we are concerned with the later stages of disk evolution, when there is no
longer any infall.  The inner regions are then supplied
only by material coming from further out in the disk plane.

In a fully turbulent disk, the surface mass density evolves with time
according to the following equation (Balbus \& Papaloizou 1999):

\begin{equation}
\frac{ \partial \Sigma}{ \partial t} = \frac{1}{r} \frac{ \partial }{
\partial r} \frac{1}{\left( r^2 \Omega \right)'} \frac{ \partial }{
\partial r} \left( \Sigma r^2  \langle W \rangle \right) ,
\label{diffusion}
\end{equation}

\noindent where $W$ is the horizontal stress tensor per unit mass
averaged over the azimuthal angle, $\langle W\rangle = \int \rho W dz / \Sigma$ is
the density--averaged of $W$ over the disk thickness and the prime
denotes derivative with respect to radius.
For an $\alpha$ disk model, $W$ is given by (Shakura \& Sunyaev 1973):

\begin{equation}
W= \frac{3}{2} \alpha c_s^2
\label{nu}
\end{equation}

\noindent where $c_s$ is the sound speed\footnote{The factor of 3/2
comes from the standard definition of $\alpha$, which involves a term
$d\ln\Omega/d\ln r$}.  There is no source term in
equation~(\ref{diffusion}) as there is no infall onto the disk.

In a layered disk, the right--hand--side of equation~(\ref{diffusion})
also gives the variation per unit time of the surface mass density of
the active zone, $\Sigma_a$, provided we replace $\Sigma$ by
$\Sigma_a$ and $\langle W\rangle$ by $\langle W\rangle_a$.  (The
notation $\langle X \rangle_a = \int_a \rho X dz / \Sigma_a$; the
integration is over the active zone).  But on the left--hand--side,
$\Sigma$ must be replaced by $\Sigma_d$, the surface density of the
dead zone.  This is because, at a given location in the disk,
$\Sigma_a$ is fixed, and determined by the X--ray source.  Therefore,
the mass which is added or removed at some radius is actually added to
or removed from the {\em dead} zone.  The evolution of the disk, in
the parts where there is a dead zone, is then formally described by
the two equations:

\begin{eqnarray}
\frac{\partial \Sigma_a}{\partial t} & = & 0 , 
\label{sigmaa} \\
\frac{ \partial \Sigma_d}{ \partial t} & = & 
\frac{1}{r} \frac{ \partial }{
\partial r} \frac{1}{\left( r^2 \Omega \right)'} \frac{ \partial }{
\partial r} \left( \Sigma_a r^2 \langle W \rangle_a \right),
\label{sigmad}
\end{eqnarray}

\noindent Note that if $\Sigma_d$ given by equation~(\ref{sigmad}) becomes
negative at some location, then it means the dead zone disappears, and
equation~(\ref{diffusion}) should be used instead to evolve $\Sigma$,
which is indistinguishable from $\Sigma_a$.
Since $\langle W \rangle_a$ does not vary with time in our models,
$\Sigma_d$ given by equation~(\ref{sigmad}) varies linearly with time.

\begin{figure}
\centerline{ \epsfig{file=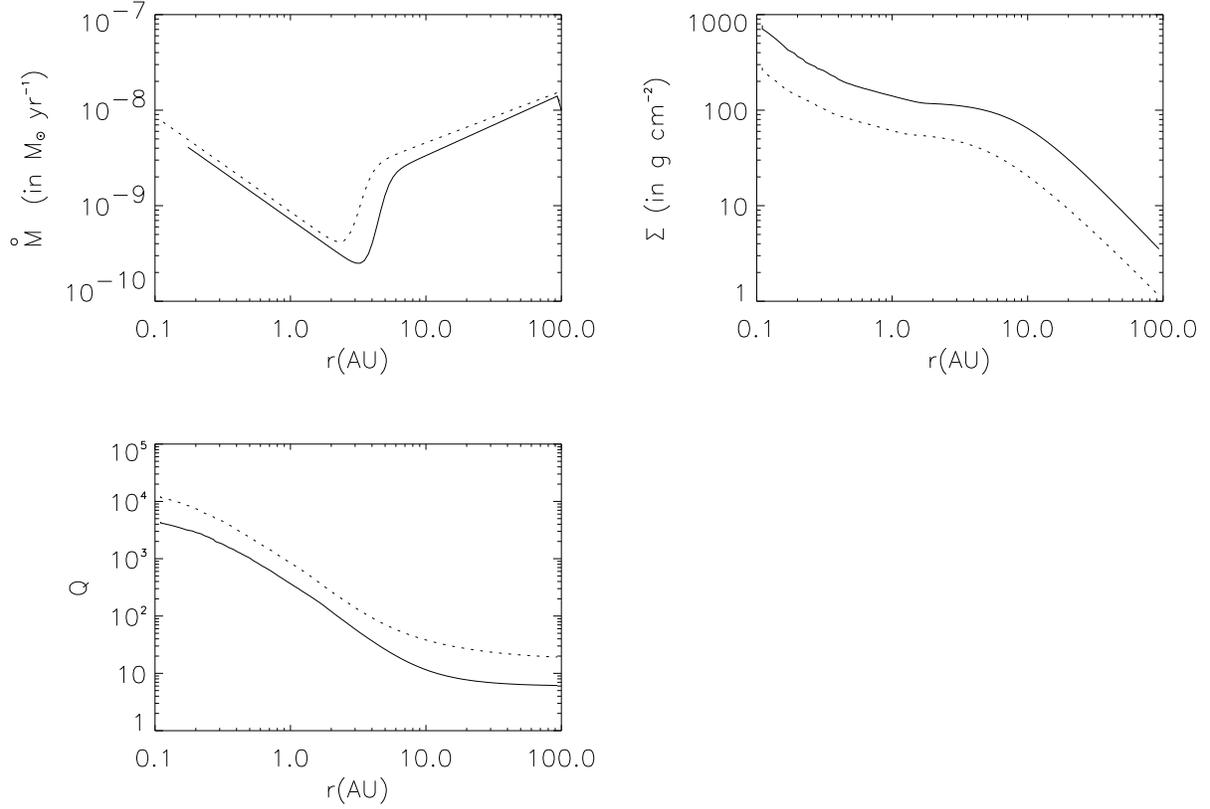,width=16.cm,angle=0}}
\caption[] {Initial conditions for the disk evolution.  We consider a
disk with $\dot{M}=10^{-8}$~M$_\odot$~yr$^{-1}$ and $\alpha=10^{-2}$
({\em solid lines}) and $3 \times 10^{-2}$ ({\em dotted lines}).
Shown are the accretion rate $\dot{M}$ through the active layer of the
disk in M$_\odot$~yr$^{-1}$ ({\em upper left panel}), the total
surface mass density $\Sigma$ in g~cm$^{-2}$ ({\em upper right panel})
and the Toomre $Q$ parameter ({\em lower left panel}) vs. $r$ in AU.
Because $\dot{M}$ increases with $r$ beyond some radius, mass is going
to accumulate in the dead zone at some intermediate radii during the
disk evolution, and $Q$ is going decrease there.  However, it does not
reach the limit for gravitational instabilities within the disk
lifetime.}
\label{fig7}
\end{figure}

We consider as initial conditions a particular disk model
characterized by the parameter $\alpha$ and a uniform $\dot{M}$.  For
illustrative purposes, we chose $\dot{M}=10^{-8}$~M$_\odot$~yr$^{-1}$
and $\alpha=3 \times 10^{-2}$ and $10^{-2}$.  The total surface mass
density of such a disk is shown in Figure~\ref{fig7}, where we also
display the Toomre parameter, $Q (r)=\Omega c_s / [\pi G \Sigma(r)]$,
where $G$ is the gravitational constant.  To get a lower limit for
$Q$, we evaluate $c_s$ at the disk surface.  Because mass is going to
accumulate at some intermediate radii in the disk, $Q$ is going to
decrease there.  We now investigate whether it can decrease enough
for gravitational instabilities to develop, i.e. for $Q$ to drop to
values $\sim 1$.  Note that for
$\dot{M}=10^{-7}$~M$_\odot$~yr$^{-1}$ and $\alpha=10^{-2}$, the
initial disk model is already gravitationally unstable beyond about
30~AU (the total mass enclosed in this radius being about
0.03~M$_{\odot}$), which is why we consider here a smaller value of
$\dot{M}$.

We compute the extent of the disk dead zone for $Re_{M, {\rm
crit}}=100$ and $x_M=0$ (see \S~\ref{sec:results}).  To discuss the
evolution of the disk, we need to consider the accretion rate through
the disk, which is obtained from (Lynden--Bell \& Pringle 1974):

\begin{equation}\label{lbp}
\dot{M} = \frac{2 \pi}{\left( r^2 \Omega \right)'} \frac{ \partial }{
\partial r} \left( \Sigma_a r^2  \langle W \rangle_a \right).
\end{equation}
(The $a$ subscripts should be dropped in the absence of a
dead zone.)  Here, $\dot{M}$ does not change with time.  In
Figure~\ref{fig7} we plot $\dot{M}$ versus $r$ at the radii where
there is a dead zone.
We see that $\dot{M}$ has a minimum at some location
$r_0$.  For $r< r_0$, $\dot{M}$ decreases with radius in the active
region of the disk.  The mass initially within $r_0$ in the dead
zone will thus gradually be accreted onto the star, and whatever mass
arrives at $r_0$ will subsequently be accreted as well.  On the
other hand, $\dot{M}$ increases with radius beyond $r_0$ in the active
layer.  Thus, in this region mass accumulates in the dead zone according to:

\begin{equation}
\Sigma \equiv \Sigma_a + \Sigma_d = \frac{1}{2 \pi r} \frac{d
\dot{M}}{d r} \; t + \Sigma_i,
\end{equation}

\noindent where $\Sigma_i$ is the initial value of 
$\Sigma$ at $r$, and equations~(\ref{sigmad}) and (\ref{lbp}) have been used.  
The evolution of the $Q$ parameter is then obtained simply from
\begin{equation}
Q\Sigma =  {\rm constant,}
\end{equation}

\noindent where the constant is simply the initial value of the
product.  For the models whose initial conditions are shown in
Figure~\ref{fig7}, $Q$ reaches values around unity first at a radius
of a few AU after a time of $2 \times 10^7$ years.  This is longer by
about an order of magnitude than the disk lifetime inferred from
observations (Haisch, Lada \& Lada 2001).  Note that this timescale is
set by the largest value of $\Sigma$ reached at the end of the
evolution and is thus insensitive to the starting conditions.  A
simple combination of layered accretion and gravitational instability
does not appear to be sufficient to truncate our fiducial disk's
lifetime rapidly enough.

\section{Discussion and conclusion}
\label{sec:discussion}

We have calculated a model for the ionization fraction of an
$\alpha$--type disk, taking into account the full vertical structure
of the disk.  In common with other investigations (Gammie 1996,
Glassgold et al. 1997), we have not included grain physics in our
ionization model, which is an important limitation.  Both the extent
and the thickness of the dead zone depend sensitively on the
parameters of the disk model, the gas accretion rate $\dot{M}$ and
$\alpha$.  For a critical Reynolds number of 100 and parameters
believed to be typical of protostellar disks, the dead zone is found
to extend from a fraction of an AU to 10--100~AU.  For
$\dot{M}=10^{-8}$~M$_\odot$~yr$^{-1}$ and $\alpha=10^{-2}$ for
instance, the dead zone extends from 0.2 to 100~AU.  For comparison,
Igea \& Glassgold (1999), who used a smaller value of $Re_{M, {\rm
crit}}$, calculated that the dead zone extended to about 5~AU.  (Also,
since in this model the mass is more centrally condensed than in ours,
the dead zone at 1~AU occupies a larger fraction of the disk vertical
column density.)  But for $\alpha=0.1$ and $\dot{M} <
10^{-7}$~M$_\odot$~yr$^{-1}$, the disk can be magnetically coupled
over its full extent.  This is because the surface mass density is low
in this case.  Models with $\alpha=10^{-3}$ have a very large and
thick dead zone.  When $Re_{M, {\rm crit}}=1$, we have found that the
dead zone is dramatically reduced or even disappears.  Clearly,
uncertainty in the input parameters profoundly influences the extent
of magnetic coupling in protostellar disks, and full coupling to the
field in some stages of disk evolution currently remains a
possibility.

We have found that the characteristics of the dead zone are
ultra--sensitive to the presence of heavy metal atoms in the disk.
This is because the metal atoms are rapidly charged by molecular ions
and recombine comparatively slowly.  Molecular dissociative
recombination is otherwise very rapid.  For $\alpha=10^{-2}$ and
$\dot{M}=10^{-8}$~M$_{\odot}$~yr$^{-1}$ for instance, an abundance of
only $10^{-6}$--$10^{-7}$ of cosmic is all that is needed in our model
for disk to be completely active, even when $Re_{M, {\rm crit}}=100$.

The density of metal atoms depends on the rate at which they are
captured by grains and the rate at which they are liberated when an
X--ray hits a grain.  (Cosmic rays, which have been ignored here, may
be important for this process.)  The resulting density is very
insensitive to the amount of dust present.  If we assume that
$\sigma_{rad} \sim \sigma_{cap}$ in equation~(\ref{nM}), then $n_M =
\epsilon L_X/(4 \pi r^2 v_{th} kT_X)$ at the distance $r$ from the
star, where $\epsilon$ is an efficiency factor (equal to the ratio of
the quantum yield to the sticking probablity which were introduced in
eq.~[\ref{nM}]).  With $L_X$ and $kT_X$ in the range
$10^{29}$--$10^{31}$~erg~s$^{-1}$ and 1--5~keV, respectively, $v_{th}
\sim 1$~km~s$^{-1}$, $\epsilon \sim 10^{-6}$ (Bringa 2001, private
communication), we get a maximum value for $n_M$ which is about
200~cm$^{-3}$ at $r=1$~AU.  For our fiducial values
$\dot{M}=10^{-8}$~M$_\odot$~yr$^{-1}$ and $\alpha=10^{-2}$, the number
density near the disk midplane at 1~AU is about $n= 6 \times
10^{13}$~cm$^{-3}$.  The maximum value of the relative density of
metal atoms is therefore $ \sim 10^{-12}$, some $10^{-7}$--$10^{-6}$
of the cosmic abundance.  This crude estimate shows that the density
of metal atoms needed for magnetic coupling is within range reached,
independently of the number of grains present.  The presence of grains
would certainly affect the ionization fraction.  This is difficult to
evaluate in a protostellar environment dominated by X-rays, and we
have not attempted the calculation in this work.  But it is important
to note that the value of $n_M$ does not depend on the density of
grains.  It remains unchanged even in a disk where {\em almost} all
the grains have sedimented toward the midplane.  This is likely to be
the situation in the later stages of disk evolution, as the
sedimentation timescale for micron--size (or larger) grains is short
(Nakagawa, Nakazawa \& Hayashi 1981).  Grain sedimentation and growth
cause characteristic changes to the radiative spectral energy
distribution.  Such changes are in fact observed, though other
interpretations cannot at present be ruled out (e.g., Beckwith,
Henning \& Nakagawa 2000).

Sano et al. (2000) investigated a finite density of 
metal atoms in a calculation of disk ionization (in their
model produced by
cosmic rays and radioactivity), together with the electrons--grains
and ions--grains reactions.  However, they assumed that the fraction
of metal atoms in the gas phase was rather high, equal to $2 \times
10^{-2}$.  In their model, the relative abundance of metal atoms is
therefore around $10^{-6}$ (they used a cosmic abundance of about $8
\times 10^{-5}$ for the refractory heavy elements), several orders of
magnitude higher than what we consider here.  Their conclusion that
the disk is almost completely turbulent when grains are sufficiently
depleted depends very much upon this assumption.

Finally, we have estimated the evolutionary timescale against
gravitational collapse of an $\alpha$ disk with a dead zone.  In our
fiducial model, we find that $\sim 10^{-3}$
M$_\odot$ accumulates between 2 and 20 AU, in regions of low optical
depth at millimeter wavelengths.  This would be easily detectable.  The
disk persists beyond $10^7$ years, an order of magnitude in excess of
observational constraints.  These results are rather robust.  The
difficulty suggests that a more elaborate model of the dead zone is
required, or that disks are more ionized than our model indicates,
or more broadly that simple layered accretion models may not
be capturing the essential dynamics of disk accretion.  A deeper
understanding of disk ionization physics clearly is essential for
further progress in this domain of star formation theory.

\section*{acknowledgment}

CT and SAB are grateful to the Virginia Institute of Theoretical
Astronomy and the Institut d'Astrophysique de Paris, respectively, for
their hospitality and visitor support.  CT acknowledges partial
support from the {\em Action Sp\'ecifique de Physique
Stellaire} and the {\em Programme National de Plan\'etologie}.  SAB is
supported by NASA grants NAG5--9266, NAG5--7500, NAG5--106555, 
and NSF grant AST 00--70979.

\newpage

\begin{table}
\begin{center}
\begin{tabular}{|c|c|c|c|c|c|c|c|}
\hline 
$Re_{M, {\rm crit}}$ & $\alpha$ & $\dot{M}$ & $x_{M,{\rm min}}$ & 
Inner radius & Outer radius & $r_{max}$ & Column density \\ 
      &          & (M$_\odot$~yr$^{-1}$)&                   & (AU)  & 
(AU)         & (AU)  &            \\
\hline 
     &         &           &            &      &    &    &      \\
1   &$10^{-2}$& $10^{-7}$ & $10^{-17}$ &  5   & 60 & 20 & 50\% \\
... &$10^{-3}$& $10^{-9}$ & $10^{-16}$ &  2   & 30 & 5  & 65\% \\
... & ...     & $10^{-8}$ & $10^{-13}$ &  0.5 & $\simeq 100$ & 10 & 90\% \\
    &         &           &            &      &    &    &      \\
100 & 0.1     & $10^{-7}$ & $10^{-14}$ &  0.4 & 30 & 10 & 55\% \\
... & ...     & $10^{-6}$ & $10^{-13}$ & 1    & $>100$ & 20 & 95\% \\ 
... &$10^{-2}$& $10^{-9}$ & $10^{-13}$ & $<0.1$ & 20 & 2  & 70\% \\
... & ...     & $10^{-8}$ & $10^{-12}$ & 0.2  & $\simeq 100$ & 6 & 90\% \\
... & ...     & $10^{-7}$ & $10^{-10}$ & 0.7 & $>100$ & 20 & 98\% \\
... &$10^{-3}$& $10^{-9}$ &            & $<0.1$ & $>100$ & 5 & 98\% \\
... & ...     & $10^{-8}$ &            & 0.4  & $>100$ & 10 & 99.9\% \\
\hline 
\end{tabular} 
\caption{Characteristics of the dead zone.  Column~1 contains $Re_{M,
{\rm crit}}$.  In columns~2 and 3 are listed the parameters which
characterized the disk, i.e.  $\alpha$ and $\dot{M}$ in
M$_\odot$~yr$^{-1}$.  Column~4 contains $x_{M,{\rm min}}$, which is
the value of $x_M$ above which there is no dead zone.  The following
columns contain the parameters which describe the dead zone for
$x_M=0$: its inner radius (which is actually insensitive to $x_M$),
its outer radius, the radius $r_{max}$ at which its vertical column
density is maximum, and the percentage of the vertical column density
it occupies at this radius.  For $Re_{M, {\rm crit}}=1$, disks with
$\alpha=0.1$ are completely active for all the values of $\dot{M}$, as
are disks with $\alpha=10^{-2}$ and $\dot{M} \le
10^{-8}$~M$_\odot$~yr$^{-1}$.  For $Re_{M, {\rm crit}}=100$, disks
with $\alpha=0.1$ and $\dot{M} \le 10^{-8}$~M$_\odot$~yr$^{-1}$ are
also completely active. For $Re_{M, {\rm crit}}=100$ and
$\alpha=10^{-3}$, there is a dead zone for all the values of $x_M$.}
\label{table1}
\end{center}
\end{table}

\newpage

\end{document}